\documentclass[twocolumn,showpacs,preprintnumbers,amsmath,amssymb]{revtex4}

\usepackage{graphicx}
\usepackage{dcolumn}
\usepackage{bm}

\begin{document}

\preprint{APS/123-QED}

\title{Resonance capture cross section of $^{207}$Pb}


\author{
\normalsize
C.~Domingo-Pardo$^{1,2}$\footnote{Corresponding author, address:
  Apdo. Correos 22085, 46071 Valencia (Spain). Tel.:+34963543499, e-mail cesar.domingo.pardo@cern.ch.},
U.~Abbondanno$^{3}$, 
G.~Aerts$^{4}$, 
H.~\'Alvarez-Pol$^{5}$, 
F.~Alvarez-Velarde$^{6}$, 
S.~Andriamonje$^{4}$, 
J.~Andrzejewski$^{7}$, 
P.~Assimakopoulos $^{8}$,    
L.~Audouin $^{2}$, 
G.~Badurek$^{9}$
% ...
\begin{center}
\normalsize The n\_TOF Collaboration\\
\end{center}
}

\affiliation{
\mbox{$^{1}$Instituto de F{\'{\i}}sica Corpuscular, CSIC-Universidad de Valencia, Spain} \\
\mbox{$^{2}$Forschungszentrum Karlsruhe GmbH (FZK), Institut f\"{u}r Kernphysik, Germany} \\  
\mbox{$^{3}$Istituto Nazionale di Fisica Nucleare, Trieste, Italy} \\ 
\mbox{$^{4}$CEA/Saclay - DSM, Gif-sur-Yvette, France} \\  
\mbox{$^{5}$Universidade de Santiago de Compostela, Spain} \\  
\mbox{$^{6}$Centro de Investigaciones Energeticas Medioambientales y Technologicas, Madrid, Spain} \\  
\mbox{$^{7}$University of Lodz, Lodz, Poland} \\ 
\mbox{$^{8}$University of Ioannina, Greece} \\  
\mbox{$^{9}$Atominstitut der \"{O}sterreichischen Universit\"{a}ten,Technische Universit\"{a}t Wien, Austria} \\ 
\mbox{$^{10}$Centre National de la Recherche Scientifique/IN2P3 - IReS, Strasbourg, France} \\  
\mbox{$^{11}$Charles University, Prague, Czech Republic} \\
\mbox{$^{12}$Dipartimento di Fisica Generale, Universit\`a di Torino, Italy} \\
\mbox{$^{13}$Universitat Politecnica de Catalunya, Barcelona, Spain} \\  
\mbox{$^{14}$International Atomic Energy Agency, NAPC-Nuclear Data Section, Vienna, Austria} \\
\mbox{$^{15}$Universidad de Sevilla, Spain} \\  
\mbox{$^{16}$Instituto Tecnol\'{o}gico e Nuclear(ITN), Lisbon, Portugal} \\  
\mbox{$^{17}$CERN, Geneva, Switzerland} \\  
\mbox{$^{18}$LIP - Coimbra \& Departamento de Fisica da Universidade de Coimbra, Portugal} \\  
\mbox{$^{19}$Istituto Nazionale di Fisica Nucleare, Bari, Italy} \\  
\mbox{$^{20}$University of Notre Dame, Notre Dame, USA} \\
\mbox{$^{21}$Centre National de la Recherche Scientifique/IN2P3 - IPN, Orsay, France} \\  
\mbox{$^{22}$Universit\`a degli Studi Pavia, Pavia, Italy} \\  
\mbox{$^{23}$Aristotle University of Thessaloniki, Greece} \\  
\mbox{$^{24}$Joint Institute for Nuclear Research, Frank Laboratory of Neutron Physics, Dubna, Russia} \\  
\mbox{$^{25}$Institute of Physics and Power Engineering, Kaluga region, Obninsk, Russia} \\  
\mbox{$^{26}$Istituto Nazionale di Fisica Nucleare(INFN), Laboratori Nazionali di Legnaro, Italy} \\
\mbox{$^{27}$Centre National de la Recherche Scientifique/IN2P3 - CENBG, Bordeaux, France} \\ 
\mbox{$^{28}$Los Alamos National Laboratory, New Mexico, USA} \\  
\mbox{$^{29}$Tokyo Institute of Technology, Tokyo, Japan} \\
\mbox{$^{30}$Oak Ridge National Laboratory, Physics Division, Oak Ridge, USA} \\    
\mbox{$^{31}$NCSR, Athens, Greece} \\
\mbox{$^{32}$Japan Atomic Energy Research Institute, Tokai-mura, Japan} \\
\mbox{$^{33}$National Technical University of Athens, Greece} \\  
\mbox{$^{34}$Institut f\"{u}r Isotopenforschung und Kernphysik, Universit\"{a}t Wien, Austria} \\
\mbox{$^{35}$P\^ole Universitaire L\'{e}onard de Vinci, Paris La D\'efense, France} \\ 
\mbox{$^{36}$CEC-JRC-IRMM, Geel, Belgium} \\
\mbox{$^{37}$Department of Physics and Astronomy - University of Basel, Basel, Switzerland} \\  
\mbox{$^{38}$ENEA, Bologna, Italy} \\  
\mbox{$^{39}$Dipartimento di Fisica, Universit\`a di Bologna, and Sezione INFN di Bologna, Italy}  
}


\date{\today}%
            
\begin{abstract} 
The radiative neutron capture cross section of $^{207}$Pb has been 
measured at the CERN neutron time of flight installation n\_TOF
using the pulse height weighting technique in the resolved energy 
region. The measurement has been performed with an optimized setup 
of two C$_6$D$_6$ scintillation detectors, which allowed us to 
reduce scattered neutron backgrounds down to a negligible level. 
Resonance parameters and radiative kernels have been determined 
for 16 resonances by means of an R-matrix analysis in the neutron 
energy range from 3~keV to 320~keV. Good agreement with previous 
measurements was found at low neutron energies, whereas substantial 
discrepancies appear beyond 45~keV. With the present results, we 
obtain an $s$-process contribution of 77$\pm$8\% to the solar 
abundance of $^{207}$Pb. This corresponds to an $r$-process
component of 23$\pm$8\%, which is important for deriving the 
U/Th ages of metal poor halo stars. 
\end{abstract}


\maketitle

\section{\label{sec:intro}Introduction}
Since $^{207}$Pb is one of the final products of the Th/U 
$\alpha$-decay chains, its abundance provides a constraint 
for the Th/U abundances and their use as cosmochronometer
\cite{cow99}. The $s$ abundance of $^{207}$Pb has been shown
to depend only weakly on details of stellar $s$-process
models \cite{rat04} so that the $r$ component can be 
reliably determined by subtraction from the solar value, 
$N_r = N_{\odot} - N_s$. 

Apart from its astrophysical importance, the neutron capture 
cross section of this isotope is also of relevance for the 
design of fast reactor systems. An eutectic mixture of Pb/Bi 
is presently considered as an appropriate spallation target 
and coolant for accelerator driven systems~\cite{rub98}.
Because 22.1\% of natural lead consists of $^{207}$Pb, the
relatively large neutron capture cross section of this 
isotope affects the neutron balance, and is therefore
important for the design of this type of hybrid reactors.

Given the large neutron scattering width in some of the 
$^{207}$Pb resonances, the neutron sensitivity of the
detector system becomes instrumental for the effective 
reduction of backgrounds due to scattered neutrons. 
For this reason an optimized setup based on C$_6$D$_6$ 
detectors has been employed by the n\_TOF collaboration,
well suited for the subsequent application of the pulse 
height weighting technique (PHWT), which was used in data 
analysis (Sec.~\ref{sec:analysis}). With the adopted 
experimental setup angular distribution effects of the 
primary $\gamma$ radiation emitted for neutron capture 
events with orbital angular momentum $l>0$ are important. 
This effect, which is large for some resonances, could be 
properly treated in the data analysis as described in 
Section~\ref{sec:analysis}. The resulting neutron capture 
cross section is presented in Section~\ref{sec:results}, 
and the astrophysical implications of this measurement 
are summarized in Section~\ref{sec:implications}.

\section{\label{sec:experiment}Experiment}

The measurement was performed with an enriched 
sample in the form of a metal disk 20~mm in 
diameter and 2~mm in thickness, containing 
92.40\% of $^{207}$Pb, 5.48\% of $^{208}$Pb and 2.12\% of $^{206}$Pb. 
The sample was mounted in vacuum inside a sample changer made from
carbon fiber together with a gold sample
for absolute cross section normalization. In addition, a 
$^{208}$Pb sample was used to study the
background coming from in-beam $\gamma$ rays, which
are scattered by the sample. The long 
flight path of 185.2~m and the short proton 
pulse width of 6 ns (RMS) are the important properties 
for achieving the high resolution in
time-of-flight (TOF) characteristic of the 
n\_TOF spallation source \cite{pro99}.

Neutron capture events were registered via the 
prompt capture $\gamma$-ray cascade by a set 
of two C$_6$D$_6$ detectors, which were optimized 
with respect to neutron sensitivity~\cite{pla03}. 
The detectors were placed at $\sim$125$^{\circ}$ 
with respect to the incident neutron beam in
order to minimize angular distribution effects of 
the emitted capture radiation. 
A schematic view of the experimental setup can be seen in  Fig.~2 of
Ref.~\cite{dom06}.
In this configuration, the in-beam
$\gamma$-ray background was also substantially reduced.

The neutron flux was monitored by means of a 
200~$\mu$g/cm$^2$ thick $^{6}$Li-foil mounted 3 m 
upstream of the $^{207}$Pb sample. Particles from  
$^{6}$Li($n, \alpha$)$^{3}$H reactions are 
registered with four silicon detectors surrounding 
the $^{6}$Li-foil outside of the beam \cite{Mar04}. 

The saturated resonance technique~\cite{mac76} using 
the 4.9~eV $^{197}$Au resonance was applied for absolute 
calibration of the $^{207}$Pb capture yield. Calibration 
of the yield in this way requires the precise knowledge of the 
neutron intensity versus the neutron energy. This has 
been determined with an uncertainty of 2\%~\cite{dom05} 
by two independent measurements performed with the 
$^{6}$Li monitor described before and with a calibrated 
fission chamber~\cite{ptb}.

\section{\label{sec:analysis}Capture data analysis}

The first aspect of the data analysis obviously concerns the 
determination of the weighting function (WF). Based on 
previous experience, the WF for the measured samples of 
$^{197}$Au (used for normalization) and of $^{207}$Pb was 
calculated via the Monte Carlo technique. The procedure 
followed has been described in detail in 
Refs.~\cite{tai02,abb04}.
 
The capture $\gamma$-ray spectra of both samples were 
calculated with a Monte Carlo code in order to estimate 
the corresponding uncertainty of the WF, which turned out to be less than
0.5\%. 
It has been experimentally demonstrated~\cite{abb04} that 
the combination of WFs obtained by the Monte Carlo method 
with the saturated resonance technique yields an overall 
systematic uncertainty of better than 3\%. However, this
level of accuracy can be only achieved if all sources of 
systematic uncertainty are properly taken into account.
This refers mainly to (i) the determination of the neutron 
flux, (ii) the treatment of the background components, and 
(iii) the effect of the electronic threshold used for 
the signals from the C$_6$D$_6$ detectors. In the particular
case of $^{207}$Pb, one has to consider also a pronounced angular
distribution effect of the capture $\gamma$ rays due to the 
low multiplicity in the de-excitation 
pattern of $^{208}$Pb. This effect will be considered 
separately since it depends strongly on the resonance spin 
and parity ($J^{\pi}$).

The various sources of systematic uncertainty are listed in
Table~\ref{tab:uncertainty}. In the following, the
treatment of these effects is described in detail.

\begin{table}[h]
\caption{\label{tab:uncertainty} Systematic effects and related 
uncertainties.}
\begin{tabular}{lc}
\hline
Effect                            & Uncertainty (in \%) \\
\hline
Weighting function, saturated resonance        &        \\
\hspace*{2mm}technique and electronic threshold & $<$2   \\

Background determination                       & $<$0.5 (1.5)$^a$  \\

Energy-dependence of neutron flux              & 2      \\

Neutron sensitivity                            & $<$0.3 \\

Angular distribution: for $J^{\pi}$=1$^{+}$    & 0.6 (8)$^b$  \\
\hspace{30mm} for $J^{\pi}$=2$^{+}$            & 0.3      \\
\hline
\end{tabular}
$^a$ Broad $s$-wave resonances at $E_{\circ} = 41$ and 256~keV (see also Table~\ref{tab:pb RP and RK}).
$^b$ Resonances with unknown angular distribution.
\end{table}

\subsection{Background}
In the present measurement the background in the 
resolved resonance region (RRR) is dominated by 
delayed $\gamma$-rays accompanying the neutron
beam and scattered in the sample. These $\gamma$-rays 
arise essentially from 
neutron capture in the water moderator of the 
n\_TOF spallation target and exhibit a smooth 
dependence on neutron time of flight. For samples of equal atomic number $Z$ and similar 
thickness the background level is also very similar. 
Since all observed resonances in $^{207}$Pb were well isolated, the
background level was best determined by choosing a relatively wide neutron energy window
around each resonance and defining the background level by a constant term.
The neutron energy dependence of the background level fitted for each
resonance was afterwards cross checked with a measurement of a $^{208}$Pb
sample. The $^{208}$Pb sample itself showed very few
resonances in the entire energy region and was, 
therefore, well suited for determination of the 
background from in-beam $\gamma$ rays. The respective
systematic uncertainties were less than 1.5\% for 
the broad resonances at 41 and 256~keV, and below 0.5\% for
all the others.

A different type of background might arise in the 
measurement of resonances that show a dominant 
scattering channel like the $s$- and $p$-wave 
resonances at 41~keV and 128~keV (Table~\ref{tab:pb 
RP and RK}), where $\Gamma_n/\Gamma_{\gamma} 
\approx 300$. Thanks to the optimized detection 
setup and the small amount of material around the 
sample and the detectors, this background turned 
out to be negligible in the present measurement.

\subsection{Electronic threshold}

The low energy cut-off in the pulse height spectra 
due to the electronic threshold ($\approx$340~keV in 
this measurement) had a non-negligible influence 
on the yield measured with the $\gamma$-ray detectors. 
The effect can be corrected by modeling the capture 
cascades with a Monte Carlo simulation of the complete 
pulse height spectra recorded in the n\_TOF detection 
system as described in Refs.~\cite{tai02,abb04,dom05}.
 
Following neutron captures on $^{207}$Pb, the 
de-excitation pattern of the $^{208}$Pb resonances is 
rather simple. It basically consists of a single 
$\gamma$-transition of 7.37~MeV for $J=1$ and of 
a two-step cascade in case of $J=2$ resonances 
\cite{ram77,ram78}. Therefore, the Monte Carlo 
simulations of the capture spectra and the estimate 
of the threshold effect are correspondingly 
straightforward and reliable, yielding correction 
factors of 5$\pm$1\% and 4$\pm$1\% for $J=1$ and $J=2$ 
resonances, respectively. 

\subsection{\label{subsec:angular}Angular 
distribution effects}

The low multiplicity ($m= 1$ for $J=1$ and $m=$ 1-2 for
$J=2$ resonances) of the $^{207}$Pb capture cascades has 
the experimental disadvantage that it causes strong
angular distribution effects in the measured pulse height 
spectra. Indeed, captures with orbital angular momenta 
$l>0$ lead to aligned states in the compound nucleus, 
perpendicular to the incident neutron beam. This causes 
anisotropy in the angular distribution of the prompt
$\gamma$ rays,

\begin{eqnarray}\label{eq:angular distribution}
W(\theta) = \sum_k A_k P_k(cos\theta) = 1 + A_2 
  P_2(cos\theta) + \nonumber\\
  + A_4 P_4(cos\theta) + A_6 P_6(cos\theta),
\end{eqnarray}

\noindent
where $P_k(cos\theta)$ are the Legendre polynomials of 
order $k$ and $A_k$ are coefficients, which depend on the 
initial ($J$) and final ($J'$) spin values, on the 
multipolarities ($L$) of the transition,
and on the degree of alignment. With ideal $\gamma$ 
detectors of negligible volume, this effect would be 
minimized by setting both detectors at 125$^{\circ}$.
However, the C$_6$D$_6$ 
detectors ($\sim$1~l volume) cover a substantial solid
angle. Hence, $\gamma$-rays are registered in a relatively 
broad angular range around 125$^{\circ}$. 

For resonances with spin $J^{\pi}=1^{+}$ there is a channel spin 
admixture ($s=0,1$), which contributes to an incomplete 
alignment with generally unknown proportions of the two 
channels with $s=0$ and $s=1$. The $A_k$ coefficients in
Eq.~\ref{eq:angular distribution} could be determined
for the two 1$^{+}$ resonances at 30.5~keV and 37.7~keV 
(Table~\ref{tab:pb RP and RK}) by means of measured 
angular distributions~\cite{bow70} and using the formulae in Ref.\cite{fer65}. With this information
the Monte Carlo simulations of the experimental setup yielded correction factors 
of  $f^{\theta,1^+}_{30~keV}=0.965(3)$ and
$f^{\theta,1^+}_{37~keV}=1.037(6)$ as described in detail 
in Ref.~\cite{dom05}. For the other three 1$^{+}$ resonances 
at 90~keV, 128~keV and 130~keV, where angular distributions
are unknown, the  corrections could not be determined. 
Instead, a systematic uncertainty of 8\% was adopted in 
these cases. This value was estimated by means of Monte Carlo simulations of
our experimental setup using different angular distributions, ranging
from alignment zero up to total alignment.

For resonances with $J^{\pi}$=2$^{+}$ the de-excitation 
occurs predominantly by emission of a two-step cascade
\cite{ram77}. The angular distribution of the first 
$\gamma$-ray can be calculated using the formulae in 
Ref.~\cite{fer65}. The angular distribution of the second 
$\gamma$-ray is influenced by the re-alignment introduced 
by the previous transition and it could be also calculated by modifying the
formulae in Ref.\cite{fer65}. The final yield correction factor for 2$^{+}$ 
resonances (considering also a 10\% branching to the ground 
state) results in a value $f^{\theta,2^+}=1.015(3)$.

\subsection{R-matrix fits}

\begin{figure}[h]
\includegraphics[width=0.45\textwidth]{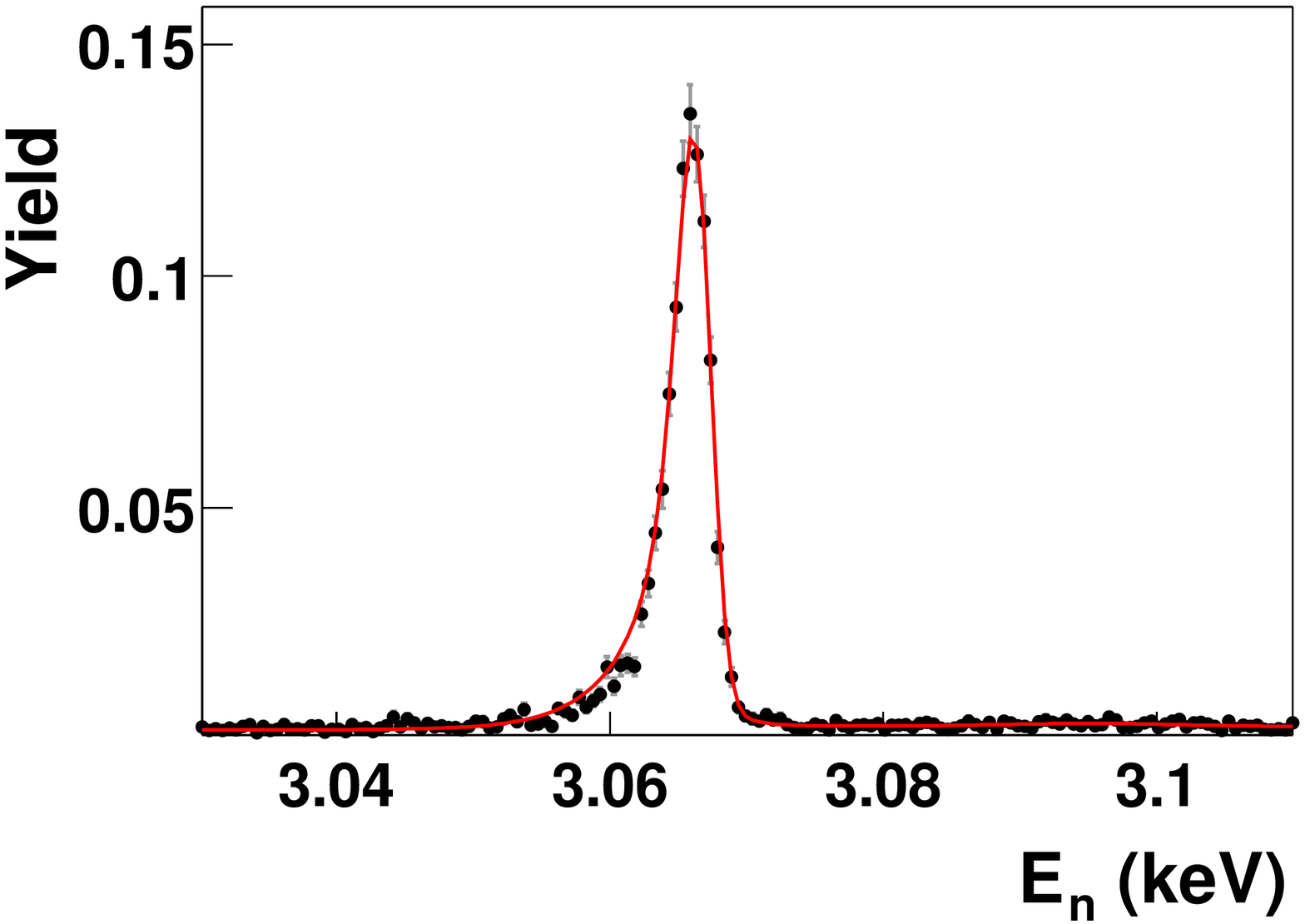}
\includegraphics[width=0.45\textwidth]{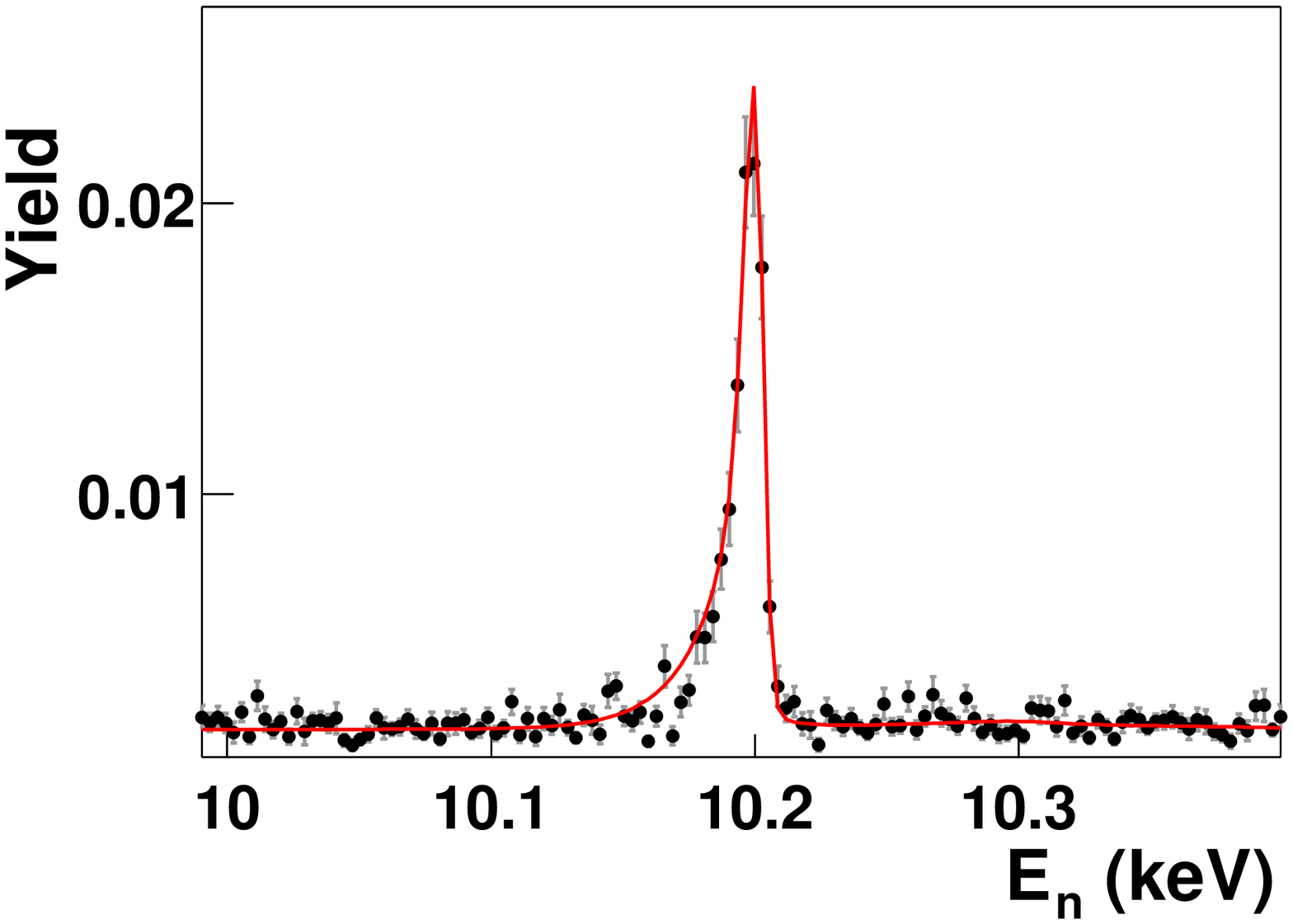}
\caption{(Color online) R-matrix fit for two of the resonances measured in
  this work at 3~keV and 10~keV.}\label{fig:resos}
\end{figure}

The capture yield corrected for the effects 
described above,
\begin{equation}\label{eq:yield}
   f^t \times f^{\theta} \times Y^{exp} = B + Y(E_{\circ},\Gamma_{\gamma},\Gamma_n),
\end{equation}
was analyzed by means of the R-matrix analysis code 
SAMMY~\cite{lar00}. $B$ is a constant term describing 
the background in the region of each resonance.
In cases, where the neutron width $\Gamma_n$ is well 
known from transmission measurements and considerably 
larger than the capture width $\Gamma_{\gamma}$, 
$\Gamma_n$ was kept fixed in our analysis. In other 
cases and where the number of counts in the
resonance was sufficiently large, we preferred to 
vary $\Gamma_n$ and $\Gamma_{\gamma}$ in order to 
better describe the corresponding capture area
or radiative kernel. Fig.~\ref{fig:resos} shows the measured capture yield
 for the two first resonances in $^{207}$Pb. The continuous line corresponds to an
R-matrix fit performed with the SAMMY code.

\section{\label{sec:results}Results and uncertainties}

The parameters and radiative kernels of the analyzed 
$^{207}$Pb resonances are summarized in 
Table~\ref{tab:pb RP and RK}. Orbital angular 
momenta $l$ and resonance spin $J$ were taken from 
Ref.~\cite{mug84}. 
\begin{table}[h]
\caption{\label{tab:pb RP and RK} Resonance parameters 
  and radiative kernels from the analysis of the 
  $^{207}$Pb(n,$\gamma$) data measured at n\_TOF$^a$.}
\begin{tabular}{cccccc}
\hline
$E_{\circ}$  & $l$ & $J$  & $\Gamma_n$ & $\Gamma_{\gamma}$ & $g\Gamma_{\gamma}\Gamma_n/\Gamma$\\
 (eV) &  &   & (meV) & (meV) & (meV)\\
\hline
\hline
3064.700(3) & 1 & 2 & 111.0(8) & 145.0(9) & 78.6(9)\\
10190.80(4) & 1 & 2 & 656(50) & 145.2(12) & 149(14)\\
16172.80(10) & 1 & 2 & 1395(126) & 275(3) & 287(30)\\
29396.1 & 1 & 2 & 16000 & 189(7) & 234(9)\\
30485.9(5) & 1 & 1 & 608(45) & 592(50) & 225(30)\\
37751(3) & 1 & 1 & 50$\times$10$^{3}$ & 843(40) & 620(30)\\
41149(46) & 0 & 1 & 1.220$\times$10$^{6}$ & 3970(160) & 2970(120)\\
48410(2) & 1 & 2 & 1000 & 230(20) & 235(20)\\
82990(12) & 1 & 2 & 29$\times$10$^{3}$ & 360(30) & 444(30)\\
90228(24) & 1 & 1 & 272$\times$10$^{3}$ & 1615(100) & 1200(80)\\
127900 & 1 & 1 & 613$\times$10$^{3}$ & 1939(150) & 1449(120) \\
130230 & 1 & 1 & 87$\times$10$^{3}$ & 900(80) & 675(60)\\
181510(6) & 0 & 1 & 57.3$\times$10$^{3}$ & 14709(500) & 8780(300)\\
254440 & 2 & 3 & 111$\times$10$^{3}$ & 1219(90) & 2110(150)\\
256430 & 0 & 1 & 1.66$\times$10$^{6}$ & 12740(380) & 9482(280)\\
317000 & 0 & 1 & 850$\times$10$^{3}$ & 10967(480) & 8120(350)\\
\hline
\end{tabular}
$^a$Orbital angular momenta $l$ and resonance spins $J$ are from 
Ref.~\cite{mug84}.
\end{table}

According to the discussion in the previous section, the 
radiative kernels can be given with an overall systematic 
uncertainty of $\approx$3\%, except for the resonances at 
90~keV, 128~keV and 130~keV with spin $J=1$, where a systematic uncertainty of 8\% had
to be adopted (see Table~\ref{tab:uncertainty}).

\begin{figure}[h]
\includegraphics[width=0.45\textwidth]{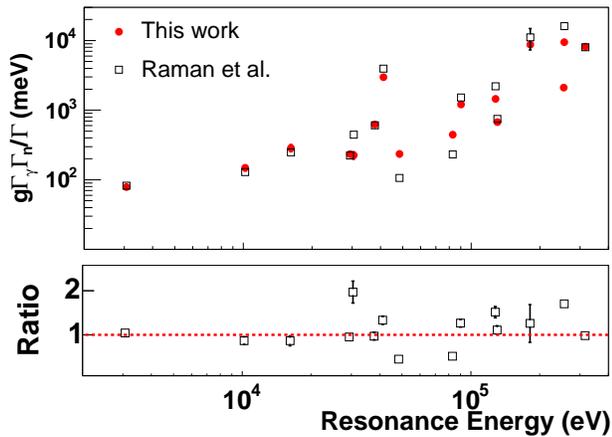}
\caption{(Color online) Comparison of the radiative capture kernels for $^{207}$Pb resonances
  measured at n\_TOF and ORNL~\cite{ram77,ram78}. In the bottom panel, the ratio between the two data sets is
shown.} \label{fig:pb RKexp}
\end{figure}

The radiative kernels (last column in Table~\ref{tab:pb RP and
RK}) are compared in Fig.~\ref{fig:pb RKexp} with previous 
measurements made at ORNL\cite{ram77,ram78}.

For the first four resonances our results are in good agreement, whereas serious
discrepancies appear at higher energy, mainly beyond 30~keV. 
It is remarkable that the radiative kernels of seven out of the 
nine $J=1$ resonances reported here are systematically smaller
than those of Ref.~\cite{ram77}, whereas the cross 
sections for $J=2$ resonances are in agreement or higher.
Such discrepancies can not be explained only in terms of 
the experimental aspects discussed in the previous section, 
but must be due to the hardness of the capture $\gamma$-ray
spectra. Since $J=1$ resonances exhibit substantially harder 
spectra, the discrepancy suggests a problem with the WF used in
the previous work. A further hint in this direction is
obtained from the fact that a thin sample (0.5~mm in thickness) 
was used in Ref.~\cite{ram77} for the neutron energy 
range below 45~keV, whereas a much thicker sample (16~mm)
was employed above that energy. The better 
agreement with the thin-sample measurement suggests that the 
WF calculated for the analysis of Ref.~\cite{ram77,ram78} 
was only valid for the thin sample case, reflecting the 
strong influence of the sample thickness on the shape of 
the WF as noted in Ref.~\cite{tai02}.

\section{\label{sec:implications}Implications for the $s$-abundances 
in the Pb-Bi region}

\begin{figure}[h]
\includegraphics[width=0.45\textwidth]{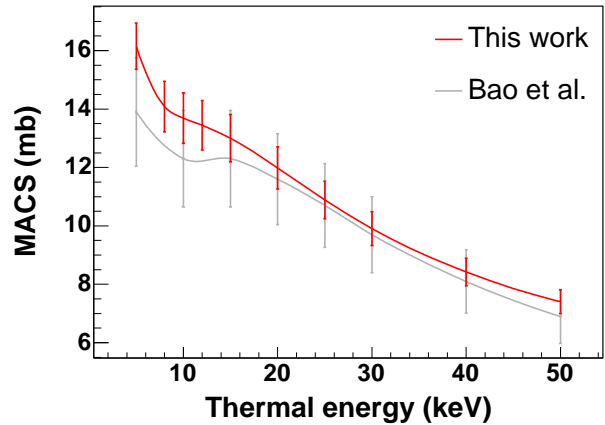}
\caption{(Color online) Maxwellian averaged cross sections for a range of 
thermal energies compared to previous values~\cite{bao00}.}
\label{fig:MACS}
\end{figure}

The Maxwellian averaged cross section (MACS) obtained 
from the present results are compared in Fig.~\ref{fig:MACS} 
with the values reported in Ref.~\cite{bao00}. Within the
quoted uncertainties, both cross sections are generally in 
good agreement. However, the new MACSs are clearly higher 
at energies below $kT$ = 15 keV. In this region, the 
uncertainties could be reduced from 13\% down to 5\%, an
improvement by more than a factor of two.

In the calculation of the MACS our results were complemented 
with some additional resonances from Ref.~\cite{suk98}, which 
could not be observed at n\_TOF due to the in-beam $\gamma$-ray
background. However, the contributions of these resonances are
rather small, i.e. only 2\% and 7\% of the  MACS at $kT=5$ and 
25~keV, respectively.

\begin{table}[h]
\caption{\label{tab:MACS}Maxwellian averaged cross 
sections  of $^{207}$Pb.}
\begin{ruledtabular}
\begin{tabular}{ccc}
$kT$  & \multicolumn{2}{c}{MACS (mb)} \\
(keV) &   Ref.~\cite{bao00}  & This work         \\
\hline
5     &  13.9 (19)               &  16.2(8)  \\
8     &                          &  14.1(8)  \\
10    & 12.3(16)                 &  13.7(9)  \\
20    & 11.5 (16)                &  12.0(7)  \\
25    &  10.7 (14)               &  10.9(6)  \\
30    & 9.7(13)                  & 9.9(6)    \\
\end{tabular}
\end{ruledtabular}
\end{table}

According to the Galactic chemical evolution (GCE) model 
described in Refs.~\cite{tra99,tra01}, the $s$-process 
abundance of $^{207}$Pb is essentially produced in 
low mass asymptotic giant branch (\textsc{agb}) stars.
In these \textsc{agb} stars energy is produced 
in a thin layer surrounding the inert C/O core. Subsequently,
extended and quiescent H-shell burning episodes are followed by
much shorter He shell flashes \cite{gal98}. In this 
model about 
95\% of the neutron exposure is due to the $^{13}$C($\alpha, 
n$)$^{16}$O reaction, which operates during the interpulse phase 
between He shell flashes at temperatures around $\sim 10^8$~K,
corresponding to a thermal energy of $kT \approx 8$~keV. 
At this stellar temperature the present MACS is about 13\% 
higher than the value reported in Ref.~\cite{bao00}. In order 
to estimate the effect of the higher cross section on the 
calculated $s$-process abundances, a model calculation
was made for thermally pulsing \textsc{agb} stars with $M = 3 
M_{\odot}$, $M = 1.5 M_{\odot}$ and a combination of metallicities: [Fe/H] =
$-$0.3 (most characteristic of the main $s$ component~\cite{arl99}),
and [Fe/H] = $-$1 (representative of the strong $s$ component~\cite{tra99,rat04}). In this way, the $s$-process
fraction of $^{207}$Pb could be determined as $N_s$ = 77(8)\%, instead of
82(18)\%~\cite{rat04}. In the present determination of $N_s$ for
$^{207}$Pb, the contributions obtained for the main and strong components were 60\% and
17\% of the solar $^{207}$Pb abundance, respectively.
In the evaluation of the
final uncertainty of the $s$-process abundance of $^{207}$Pb, the estimated contribution due to
the uncertainty in its cross section is now small, of about 3\%. 
The contribution due to the uncertainty in the neutron capture cross sections of $^{204}$Pb,
$^{205}$Pb, and $^{206}$Pb~\cite{bao00} has been estimated as less of 2\%. 
The uncertainties of the $s$-process model are estimated to be $\pm$3\% for the
main and $\pm$10\% for the strong component, resulting a contribution of less
than 4\% to the total uncertainty in the $s$-process abundance of $^{207}$Pb.
Finally, the main source of uncertainty in the determination of $N_s$ for $^{207}$Pb
is due to the uncertainty in the solar abundance of lead, which is of 8-10\%~\cite{and89,lod03,asp05}.

Since the $s$ abundance of this isotope is robust with respect to uncertainties in the
$s$-process model, the $r$ component can now be constrained as
$N_r=23(8)$\%. 
Within the quoted uncertainty of 35\%, this result is compatible with 
earlier $r$-process calculations~\cite{cow99} based on the
ETFSI-Q nuclear mass model, which find an $r$-process 
fraction for $^{207}$Pb of 18.4\% with an uncertainty of 10-20\%,
and with  $r$ abundance calculations reported
more recently~\cite{kra04}, which yield values between 15.1\% and 16.4\%.

\section{Conclusions}

The neutron capture cross section of $^{207}$Pb has been
measured at the high resolution neutron time-of-flight facility 
n\_TOF.

The main sources of systematic uncertainty affecting this 
measurement have been treated in detail. From the experimental 
point of view background due to scattered neutrons could be 
eliminated by means of an optimized detection setup, and 
angular distribution effects were minimized by setting the 
detectors at 125$^{\circ}$ with respect to the neutron beam. 
In data analysis, the remaining systematic effects have 
been carefully corrected, resulting in a 3\% accuracy for 
the radiative kernels of the capture resonances, except 
for three cases, where missing angular distribution data led 
to an uncertainty of 8\%.

We report resonance parameters and capture areas for 16 
resonances in the neutron energy interval from 3~keV to 
320~keV. At low neutron energies our results are in
good agreement with previous data, but reveal significant
discrepancies for resonances above 40~keV. Maxwellian 
averaged cross sections for $^{207}$Pb were determined 
with an accuracy of $\pm$5\%. With this information
the $s$-process component of solar $^{207}$Pb was determined
to be 77(8)\% resulting in an $r$ fraction of 23(8)\%.

\begin{acknowledgments}
  This work was supported by the EC under contract FIKW-CT-2000-00107,
  by the Spanish Ministry of Science and Technology (FPA2001-0144-C05)
  and by the funding agencies of the participating institutes. It is part of
  the PhD-Thesis of C.D. who acknowledges financial support from Consejo
  Superior de Investigaciones Cient{\'{\i}}ficas.

\end{acknowledgments}

\newpage 
\bibliography{DomingoPRC}
\end{document}